\documentclass[conference]{IEEEtran}
\usepackage{graphicx}
\usepackage{subfigure}
\usepackage{float}
\usepackage{mathrsfs}
\usepackage{amsfonts}
\usepackage{array}
\usepackage{amssymb,amsmath}
\usepackage{longtable}
\usepackage{rotating}
\usepackage{multirow}
\usepackage{epstopdf}
\usepackage[lined,boxed,commentsnumbered, ruled]{algorithm2e}
\usepackage{cite}
\usepackage{float}
\usepackage{stfloats}
\usepackage[colorlinks,
            linkcolor=black,
            anchorcolor=black,
            urlcolor=black,
            citecolor=black]{hyperref}
\usepackage[hyphenbreaks]{breakurl}

\renewcommand{\baselinestretch}{1}

\def\BibTeX{{\rm B\kern-.05em{\sc i\kern-.025em b}\kern-.08em
    T\kern-.1667em\lower.7ex\hbox{E}\kern-.125emX}}

\renewcommand{\baselinestretch}{1}
\allowdisplaybreaks[4]
\newcommand{\systemname}{ISAC}

\begin{document}
\title{Optimizing Radio Access Technology Selection and Precoding in CV-Aided ISAC Systems}
\author{\IEEEauthorblockN{Yulan Gao\textsuperscript{1}, Ziqiang Ye\textsuperscript{2}, Ming Xiao\textsuperscript{1} 
and Yue Xiao\textsuperscript{2} }
\IEEEauthorblockA{\textsuperscript{1}{Division of Information Science and Engineering} \\
{KTH Royal Institute of Technology, 100 44 Stockholm, Sweden}\\
\IEEEauthorblockA{\textsuperscript{2}{National Key Laboratory of Wireless Communications}\\
{University of Electronic Science and Technology of China, Chengdu, 611731, China}}
{Email: yulang@kth.se, yysxiaoyu@hotmail.com, mingx@kth.se, xiaoyue@uestc.edu.cn}}
}

\maketitle
\begin{abstract}
Integrated Sensing and Communication (\systemname{}) systems promise to revolutionize wireless networks by concurrently supporting high-resolution sensing and high-performance communication.
This paper presents a novel radio access technology (RAT) selection framework that capitalizes on vision sensing from base station (BS) cameras to optimize both communication and perception capabilities within the \systemname{} system.
Our framework strategically employs two distinct RATs, LTE and millimeter wave (mmWave), to enhance system performance.
We propose a vision-based user localization method that employs a 3D detection technique to capture the spatial distribution of users within the surrounding environment. 
This is followed by geometric calculations to accurately determine the state of mmWave communication links between the BS and individual users. 
Additionally, we integrate the SlowFast model to recognize user activities, facilitating adaptive transmission rate allocation based on observed behaviors.
We develop a Deep Deterministic Policy Gradient (DDPG)-based algorithm, utilizing the joint distribution of users and their activities, designed to maximize the total transmission rate for all users through joint RAT selection and precoding optimization, while adhering to constraints on sensing mutual information and minimum transmission rates.
Numerical simulation results demonstrate the effectiveness of the proposed framework in dynamically adjusting resource allocation, ensuring high-quality communication under challenging conditions.
\end{abstract}


\section{Introduction}\label{sec:intro}

The landscape of wireless service applications, such as autonomous vehicles, holography-based communications, and immersive virtual/augmented reality, is rapidly evolving, requiring a more integrated and flexible approach within wireless networks.
These advanced applications demand not only reliable communication but also precise environmental sensing, with real-time processing and minimal delay \cite{zhang20196g}.
While current the fifth-generation (5G) networks provide enhanced broadband services, their capabilities in environmental sensing are limited, particularly when line-of-sight communication is obstructed.
The clear separation of sensing and communication in 5G networks is insufficient to meet the complex requirements of next-generation applications \cite{mao2017survey}.
To overcome these limitations, Integrated Sensing and Communication (ISAC) has emerged as a new paradigm aimed at optimizing resource allocation to address the needs of advanced applications in forthcoming the sixth-generation (6G) networks \cite{feng2021joint}.

One key challenge occurs when obstacles prevent direct communication between the base station (BS) and users in millimeter-wave (mmWave) communication scenarios.
In response, we propose a novel framework that leverages cameras to assist in locating and communicating with users obstructed by obstacles.
The cameras, installed above the BS, can detect the location and distance of users relative to the BS, even when direct mmWave communication is not possible.
With this information, the BS switches to an alternative LTE link \cite{khawam2016radio}, ensuring continuous connectivity and maintaining service even when mmWave signals are blocked.
Moreover, the visual data captured by the cameras enables further functionality.
Using the SlowFast model \cite{fan2020pyslowfast}, the system can analyze the behavior of users from the images.
The behavior recognition results are then used to determine the minimum transmission rate for each user \cite{zhang2020device,gao2017csi}, ensuring that their communication needs are met based on their activity.

In this paper, we aim to maximize the overall transmission rate for all users, ensuring that the sensing mutual information (MI) meets minimum requirements \cite{ni2021multi,yang2007mimo} and that each user’s transmission rate is appropriately aligned with their observed behavior.
This ensures efficient communication without sacrificing the sensing accuracy required for location and behavior detection.
The detailed contributions of this work are summarized as follows:
\begin{itemize}
    \item We introduces a novel RAT selection framework that utilizes vision sensing from BS cameras to optimize communication and perception capabilities in the \systemname{} system, employing two distinct RATs: LTE and mmWave.
    We use the 3D objective detection technique \cite{qin2019triangulation} to extract spatial distribution of users from images captured by BS cameras.
    \item Subsequently, we utilize the SlowFast model to analyze visual data and accurately identify user behaviors, thereby enabling precise adjustment of transmission rates tailored to individual activities.
    \item Based on the above analysis, we develop a Deep Deterministic Policy Gradient (DDPG)-based algorithm that dynamically optimizes RAT selection and precoding. This algorithm aims to maximize the overall transmission rate while ensuring the minimum sensing MI requirements and meeting individual user transmission needs.

\end{itemize}

The remainder of this paper is structured as follows: Section II presents the system model and problem formulation. Section III outlines the proposed framework and algorithm design. Section IV provides detailed simulation results to validate the performance of the framework, and Section V concludes the paper.

\section{System Model and Problem Formulation}
\subsection{System Overview}
\begin{figure}[!t]
\centering
\includegraphics[width=\linewidth]{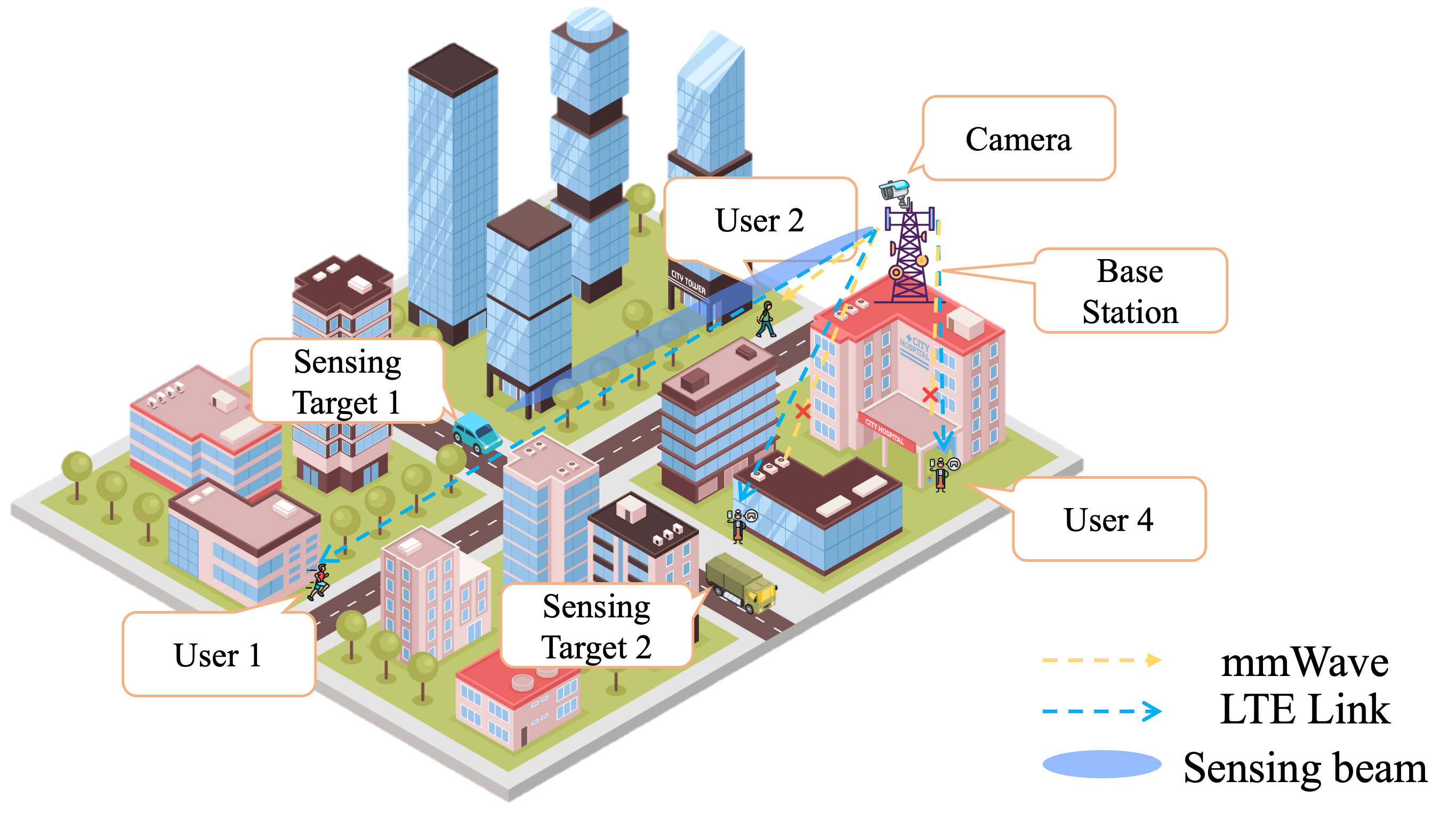}
\caption{The considered ISAC system with vision sensing.}
\label{fig:2}
\end{figure}

Consider an \systemname{} system that incorporates a widespread network of visual sensing devices, as depicted in Fig. \ref{fig:2}. 
These devices are deployed across a broad spectrum of public spaces, encompassing educational settings, parks, healthcare facilities, and recreational areas, among others.

Consider an area serviced by a BS equipped with $M$ antennas and several cameras, supporting $N$ single-antenna users, denoted by the set $\mathcal{N} = \{1, 2, \ldots, N\}$, and $L$ far-field sensing targets represented as $\mathcal{L} = \{1, 2, \ldots, L\}$.
Building on \cite{hijazi2016machine}, we assume that each user is doing an activity, with the entire scope pf possible activities are contained within a set $\mathcal{G}=\{1,2,\ldots,G\}$.
Let $\mathbf a=\{a_n|a_n\in \mathcal{G}\}_{n=1}^{N}$ denote the activity profile of all users.
To successfully communicate with the BS, each user $n$ should first select its RAT.
To represent these selections, we introduce a binary variable $\boldsymbol x = \{x_n, \forall n \in \mathcal{N}\}$, defined as follows:
\begin{align}
    x_n=\left\{\begin{matrix}
1,  & \text{if the link between user }n \text{ and BS via mmWave;} \\
0,  & \text{otherwise}.
\end{matrix}\right.
\end{align}

\subsection{Communication Model}

Following \cite{jiang2022rethinking}, the achieved data rate by mobile user $n$ under polite resource reuse takes the form:
\begin{align}\label{eq:8}
&R_n^{\mathrm{mm}}(\gamma_n)=\frac{B_n\left(L-L_{p}\right)}{\left[1-1 /\left(1+\sigma_{1}^{2} \gamma_{n} L_{p}\right)\right] \ln 2}\\ \notag
&\left[e^{\frac{1+\gamma_{n} L_{p}}{\gamma_{n}}} \operatorname{Ei}\left(-\frac{1+\gamma_{n} L_{p}}{\gamma_{n}}\right)-\right.
\left.e^{\frac{1}{\gamma_{n} \sigma_{1}^{2}}} \operatorname{Ei}\left(-\frac{1}{\gamma_{n} \sigma_{1}^{2}}\right)\right], 
\end{align}
where $B_n$ denotes the allocated bandwidth, 
$L$ is the total number of symbols in one time slot,
$L_p$ represents the number of pilot symbols, $\sigma_1^2$ shows the variance of the communication channel, $\operatorname{Ei}(\cdot)$ is the exponential integral function, and the term $\gamma_n$ represents the Signal to Interference plus Noise Ratio (SINR) on device $n$, which can be written as 
\begin{align}\label{eq:9}
    \gamma_{n}=&10\log_{10}\left[\frac{|{\boldsymbol H}_n^H{\boldsymbol w}_n|^2}{\sum_{i\in \mathcal{N},x_i=x_n, i\neq n}|{\boldsymbol H}_n^H{\boldsymbol w}_i|+\sigma^2}\right]\\ \notag
    &-20\log_{10}\left[\frac{\lambda}{4\pi d_0}\right]+10\varrho\log_{10}\left [ \frac{d_n}{d_0}\right],
\end{align}
where ${\boldsymbol W}=({\boldsymbol w}_1, {\boldsymbol w}_2, \ldots, {\boldsymbol w}_N)^H\in {\mathbb C}^{N\times M}$ is the precoding matrix, $\varrho$ is the path loss index, $d_0$ is the reference distance, $d_n$ is the distance between the user $n$ and BS which can be obtained by the radar or the visual camera introduced in \ref{chap:distance_cal}, and ${\boldsymbol H}_n$ is the normalized narrow-band mmWave channel with $\mathcal{L}$ scattering paths in our multi-user multi-input-multi-output (MIMO) scenario and it can be written as
\begin{align}\label{eq:10}
{\boldsymbol H}_n=\sqrt{\frac{M}{{L}}}\sum_{l=1}^{{L}}\alpha_l {\boldsymbol a}(\phi_l),
\end{align}
where $\alpha_l$ represents the fading coefficient of the $l$-th path, which is generally modeled as a Gaussian distribution.
$\phi_l$ is the Angle of Departure (AoD) of the $l$-th path.
And $\boldsymbol{a(\phi_l)}$ represents the steering vector of the $l$-th path transmitting side
\begin{align}\label{eq:11}
  \boldsymbol{a}(\phi_l)=\sqrt{\frac{1}{M}}\left[1,e^{j\frac{2\pi}{\lambda}d\sin(\phi_l)},\ldots,e^{j\frac{2\pi}{\lambda}d(M-1)\sin(\phi_l)}\right]^T,
\end{align}
where $d$ is the antenna spacing and $\lambda$ is the wavelength.

For LTE link, the transmission rate for user $n$ when $x_n=0$ is denoted as:
\begin{align}
    R_n^{\mathrm{lte}}=B_n\log_2(1+\gamma_n).
\end{align}
Hence, the achievable sum rate for all users in set ${\mathcal N}$ is obtained as 
\begin{align}
C=\sum_{i\in{\mathcal N},x_i=1}R_i^{\mathrm{mm}}(\gamma_i) + \sum_{j\in{\mathcal N},x_j=0}R_j^{\mathrm{lte}}(\gamma_j).
\end{align}
\subsection{Problem Formulation}
Following \cite{wei2024precoding}, we introduce the sensing MI to present the amount of information in the sensing channel obtained from the ISAC waveform received at the BS. The detailed expression is shown as follows:
\begin{align}
    \text{MI}=\log_2\left[(\delta_r^2)^{S-2M} \det\left( S\boldsymbol{R_x}\Sigma_{G}+\delta_r^2 \boldsymbol{I}_{M}\right) \right],
\end{align}
where $\boldsymbol{R_x}=\boldsymbol{W}^{\mathrm{mm}}{\boldsymbol{W}^{\mathrm{mm}}}^H$, $\boldsymbol{W}^{\mathrm{mm}} = [\boldsymbol{w}_{i_1},\boldsymbol{w}_{i_2}, \ldots, \boldsymbol{w}_{i_{|\sum_{j\in{\mathcal N}}x_j|}}]^H$ represents the splicing of vectors with all $x_{i_k}=1, i_k\in {\mathcal N}$.  
$\delta_r$ is the variance of AWGN, $S$ is the number of OFDM symbols, and $\Sigma_{G}$ is the normalized sensing channel.

Given the above discussions, we introduce the joint RAT selection and precoding optimization problem as follows: 
\begin{align}
    &\max_{\boldsymbol{W, x}} \sum_{i\in{\mathcal N},x_i=1}R_i^{\mathrm{mm}} + \sum_{j\in{\mathcal N},x_j=0}R_j^{\mathrm{lte}}\label{eq:18}\\
    \text{s.t.~}&R_n \ge R_{n,t}^{\min},\forall n \in \mathcal{N},\tag{\ref{eq:18}a}\\
    &||\boldsymbol{W}||_F^2 \le p^{\max},\tag{\ref{eq:18}b}\\
    &\text{MI}\ge \text{MI}^{\min},\tag{\ref{eq:18}c}
\end{align}
where $p^{\max}$ and $R_{n,t}^{\min}$ represent the maximum allowable transmit power of the BS and the minimum data rate required for transmission based on the user's activity at slot $t$, respectively.
And $\text{MI}^{\min}$ is the predefined MI threshold for sensing.

The \systemname{}'s  objective is to optimize the beamforming ${\boldsymbol W}$ and RAT selection $\boldsymbol x$ to maximize the sum of transmission rates for all users.
Note, however, the information regarding the minimum possible rate ${R_{n,t}^{\min}} \forall n\in{\mathcal N}$ is uncertain, as it varies over time depending on the user’s activity.
Without loss of generality, for each user $n\in{\mathcal N}$, we assume that $R_{n,t}^{\min}$ is strongly related to the user's instantaneous activity, $a_n$. 
Thus, the \systemname{} should ensure continuous activity recognition for all users. 

To address this challenge, we propose a novel framework that leverages camera-based user localization and the SlowFast model for real-time activity recognition.
This framework dynamically adjusts RAT selection between mmWave and LTE, depending on the user’s environment and behavior, ensuring continuous connectivity and meeting transmission rate requirements.
Thus, in Section III, the joint RAT selection and ISAC precoding optimization algorithm is developed based on the extracted spatial distribution and user behavior information from images captured by BS cameras.  


\section{CV-Aided ISAC Framework and Algorithm Design}

\subsection{Spatial Information Calculation} \label{chap:distance_cal}

\begin{figure}[!t]
\centering
\includegraphics[width=\linewidth]{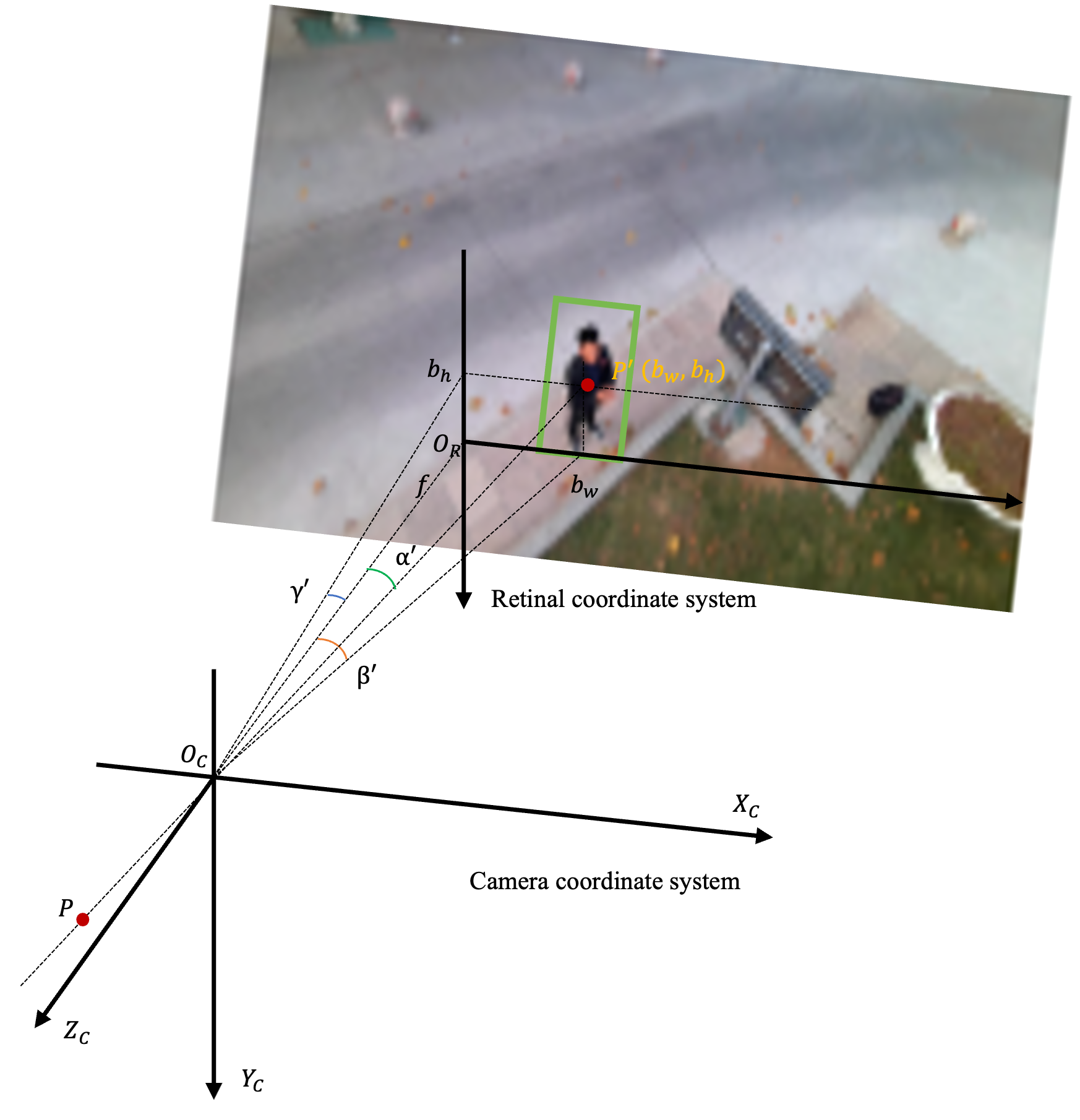}
\caption{Spatial relationship between camera and retinal coordinate system, the angle between target and its three axes is $\alpha^{\prime},\beta^{\prime}$ and $\gamma^{\prime}$.}
\label{fig:5}
\end{figure}
To optimize the precoding matrix effectively, it is crucial to determine the distance between the user and the BS.
When there are no obstacles between the user and the BS (i.e., the user is in line-of-sight (LoS) with the BS) and within the coverage of area of mmWave, the distance and angle can be directly measured via mmWave.
On the other hand, if the target cannot be detected by mmWave--either due to physical obstructions or excessive distance--computer vision techniques are required to calculate the user-to-BS distance.
Utilizing the SlowFast framework, the class and position of the user are extracted from the image.
The output includes activity classifications of the user along with bounding box data, as shown in Fig.\ref{fig:5}. Specifically, the bounding box is represented by $[b_x, b_y, b_h, b_w]$ and an 8-D offset vector, where $[b_x, b_y]$ signifies the center of the bounding box, $b_h$ denotes the bounding box height, $b_w$ refers to its width, and the 8-D vector provides offsets for the eight horizontal and vertical coordinates of the four corner points.

Relative spatial relationships include both directional and distance information \cite{geiger2012automatic}.
In the context of a camera scene, when the distance from the target to the camera (Z) significantly exceeds the focal length ($fl$) of the camera, the imaging system can be approximated by a pinhole camera model.
To map the target information from 3-D space onto 2-D image coordinates using a rigid body model transformation, the system requires four coordinate frames and a translation matrix $\boldsymbol{T}$ \cite{geiger2012automatic}.
The translation matrix $\boldsymbol{T}$ can be obtained by
\begin{align}
    \boldsymbol{T} = \begin{bmatrix}
 \boldsymbol {R} & \boldsymbol t\\
  0^T & 1
\end{bmatrix}.
\end{align}
In a static scene, the translation vector $\boldsymbol{t}$ equals zero. 
The rotation matrix $\boldsymbol{R}$ can be derived from the camera's external parameters as
\begin{align}
    \boldsymbol R = \begin{bmatrix}
 1,0,0 & 0,\cos \Psi,\sin \Psi & 0,-\sin \Psi,\cos \Psi\\
 \cos \phi,0,-\sin \phi & 0,1,0 & \sin \phi,0,\cos \phi\\
 \cos \theta, \sin \theta, 0 & -\sin \theta,\cos \theta,0 & 0,0,1
\end{bmatrix}.
\end{align}
In this context, the roll angle ($\Psi$) represents the angle between the camera and the $x$-axis of the world coordinate system, the pitch angle ($\phi$) denotes the angle between the camera and the $y$-axis of the world coordinate system, and the yaw angle ($\theta$) describes the angle between the camera and the $z$-axis of the world coordinate system.

Thus, the relationship between the camera coordinates and world coordinates can be established.
By applying the projection relationship, the target width in world coordinates is $X$ and its height is $Y$, while in the image plane, the width is denoted by $X^{\prime}=b_w$, and the height by $Y^{\prime}=b_h$.
The ratio involving the distance between the target and the optical center can therefore be transformed as follows
\begin{align}
    S=\frac{Z}{fl}=\frac{X}{X^{\prime}}=\frac{Y}{Y^{\prime}}.
\end{align}
The distance $d_n$ between the target center and the camera's focal point is calculated as $d_n = Y \times fl / Y^{\prime}$.

Additionally, the relative orientation between the transmitter and the receiver can be described by three angles: $\alpha$, $\beta$, and $\gamma$.
These angles are determined based on the center coordinates $[b_x, b_y]$ within the image coordinate system.
Specifically, $\alpha^{\prime}$ represents the angle between the $Z_C$ axis and the line extending from $O_c$ to $[b_x, b_y]$.
Meanwhile, $\beta^{\prime}$ and $\gamma^{\prime}$ denote the angles formed between the $Z_C$ axis and the corresponding line projections.
According to the pinhole imaging model, the angles $\alpha$, $\beta$, and $\gamma$--which describe the orientation between the target center and the three axes of the camera coordinate system--are equivalent to those between the image center and the coordinate axes.
\begin{align}
    \left[\begin{array}{l}
\alpha^{\prime} \\
\beta^{\prime} \\
\gamma^{\prime}
\end{array}\right]=\left[\begin{array}{l}
\alpha \\
\beta \\
\gamma
\end{array}\right]=\left[\begin{array}{c}
\arctan \frac{b_{x}}{f} \\
\arctan \frac{b_{y}}{f} \\
\arctan \frac{\sqrt{b_{x}^{2}+b_{y}^{2}}}{f}
\end{array}\right].
\end{align}

\subsection{Algorithm Design}
Building on the above-designed framework, we employ the DDPG algorithm to address the optimization problem \eqref{eq:18}. 
DDPG, a model-free, off-policy actor-critic method integrated into a deep learning framework, is particularly suited for continuous action spaces, making it ideal for optimizing high-dimensional, continuous action domains such as those found in dynamic RAT selection and precoding design.
This approach ensures optimization of both precoding matrix and RAT selection at BS. 
We denote the state space of problem \eqref{eq:18} at time slot $t$ as ${\mathbb S}=(\{\boldsymbol H_{n,t}\}_{n\in{\mathcal N}}, {\boldsymbol a}_t, \{d_{n,t}\}_{n\in{\mathcal N}}, \{R_{n,t}^{\min}\}_{n\in{\mathcal N}})$. 
It encompasses the essential aspects that influence the performance of the precoding matrix in an \systemname{} system.
The action space is represented by the RAT selection matrix $\boldsymbol x$ and the precoding matrix $\boldsymbol{W}$, a continuous and multidimensional entity that directly influences the beamforming strategy and signal quality for users.
The corresponding reward is set as \eqref{eq:18}. 
To emphasize the importance of adhering to the system's constraints, any action that violates these constraints incurs a significant penalty, with the reward set to $-100$.
More details of the joint RAT selection and ISAC precoding optimization algorithm is shown in Algorithm \ref{alg:1}.

\begin{algorithm}[!t]
\SetAlgoLined 
\caption{Joint RAT Selection and ISAC Precoding Optimization Algorithm}\label{algorithm:1} \label{alg:1}
{\bf Require:} $p^{\max}$, $\{R_n^{\min}\}_{n=1}^N$, $MI^{\min}$\\
\KwIn{Initialize the SlowFast model for activity detection;\
Initialize DRL model for adaptive precoding optimization;
}
\eIf{User $n$ is in LoS with the BS and within the coverage area of mmWave}
{Estimate the distance $d_n$ based on the mmWave radar.\\}
{Estimate the distance $d_n$ based on the visual camera.\\}
Estimate the channel $\boldsymbol{H}_n$ following Eq. \eqref{eq:10}.\\
Capture raw visual data from the vision monitor.\\
Use SlowFast model to detect current activity.\\
\eIf{activity detected}
{Store the activity results;\\
}
{Continue to the next iteration.}
Send state space to the DRL model to get ${\boldsymbol W}=({\boldsymbol w}_1, {\boldsymbol w}_2, \ldots, {\boldsymbol w}_N)^H\in{\mathcal C}^{N\times M}$ and $\boldsymbol{x}$ to maximize the reward given in Eq. \eqref{eq:18};

\KwOut{Dynamically precoding matrix and RAT selection based on user activities for maximizing utility.}
\end{algorithm}

\section{Simulation Results}
\begin{table}[!t]
\centering
\caption{Hyperparameters Used in DDPG}
\begin{tabular}{c|c|c}
  \hline
   {\bf Parameters} & {\bf Symbol} & {\bf Value} \\
  \hline
   Learning rate for actor network& $\eta_a$ & $0.001$\\
 
  Learning rate for critic network& $\eta_c$ & $0.001$\\

 Discount factor & $\gamma$ & $0.99$\\
  
   Soft update parameter & $\tau$ &  $0.005$\\
  
   Replay buffer size & $RB$ & $10000$\\
  
   Batch size & $BS$ & $64$\\
  
 Max step & $ES$& $3000$\\
 
   Explore noise & $noise$ & $0.2$\\
  \hline
\end{tabular}
\label{table:1}
\end{table}
\subsection{Simulation Environments and Settings}
In this paper, we delineate a comprehensive spectrum of human physical activities, categorizing them into a set of $60$ distinct types.
The {\tt UAV-Human} dataset \cite{Li_2021_CVPR} focuses on spatiotemporal localization of human actions.
The UAV-Human dataset contains $155$ action classes, which ranges from sit down, kick something, make a phone call to play with cell phones.

The wavelength of mmWave, $\lambda_{\mathrm{mm}}$, is established at $2\mbox{ mm}$, the LTE wavelength, $\lambda_{\mathrm{LTE}}$ is set at $0.1\mbox{ m}$, and the number of pilot symbols, $L$, is fixed at $14$.
mmWave and LTE have $5$ and $9$ scattering paths, respectively.
If not otherwise specified, the number of users is $10$ and the number of antennas is $16$.
The lower bound of sensing MI, $\mbox{MI}^{\min}$, is $90\mbox{ bits}$.
The symbol time interval $T_s$ is set to $0.05$ ms, and the radar cross-sectional area $s_{\text{rcs}}$ is $100~\text{m}^2$. 
The variance of communication channel $\sigma_1^2=2$, and the Rice factor of perception channel $K=A_s/\sigma_2^2=3$ while the root-mean-square bandwith $B_{\text{rms}}=\sqrt{12}B_n$.

\begin{figure}[!t]
\centering
\includegraphics[width=\linewidth]{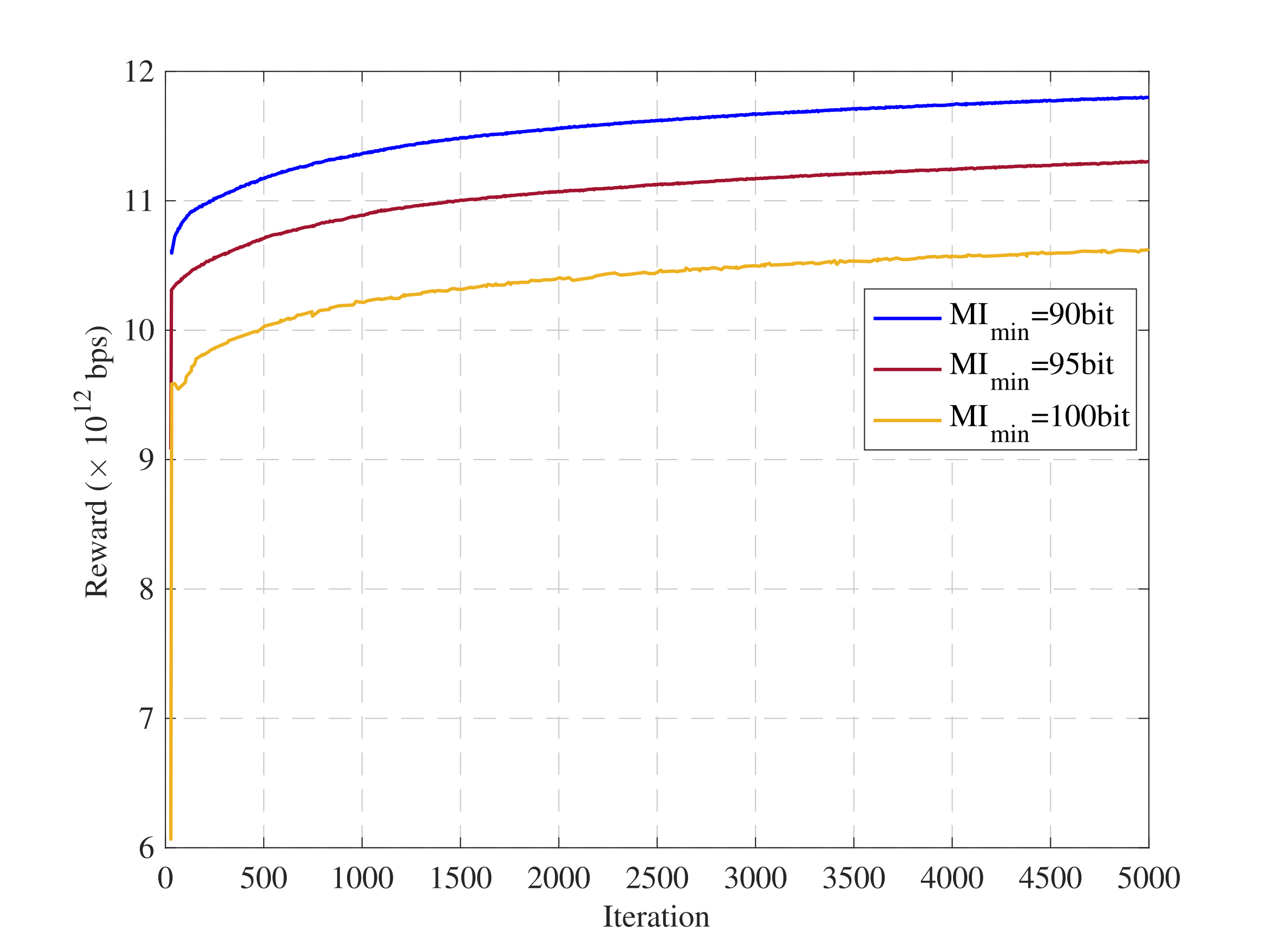}
\caption{Convergence of Reward under Different Minimum Mutual Information (MI) Constraints.}
\label{fig:3}
\end{figure}

The Actor network consists of two fully connected layers, where it processes the input state and outputs an action using Rectified Linear Unit (ReLU) and Sigmoid activation functions respectively, with the Sigmoid ensuring that the action values are within a specified range.
On the other hand, the Critic network, also comprising two fully connected layers, takes both the state and action as inputs, merges them, and then outputs a single value representing the estimated value of the state-action pair, using a ReLU activation function in its first layer.
The learning rate is set to $0.001$, discount factor is set to $0.99$ and the variance of explore noise is $0.2$.
The soft update factor is $0.005$.
The batch size is set to $64$ and the memory buffer size is $10000$.
To clearly delineate the parameters of the DDPG algorithm, we have enumerated the hyperparameters in Table \ref{table:1}.

In order to gain insight into the proposed joint RAT selection and ISAC precoding optimization algorithm, we select the following three relevant baselines for assessment.   
\begin{itemize}
\item {\bf\em Random RAT selection}: RAT is randomly selected.
\item {\bf\em mmWave}: All users are forced to access mmWave. This is the typical ISAC system setup.
\item {\bf\em LTE}: all users are restricted to using LTE, resulting in zero sensing mutual information.
\end{itemize}

\subsection{Results and Discussion} 

Figure \ref{fig:3} demonstrates the convergence of the reward over iterations for three distinct minimum MI requirements ($\mbox{MI}^{\min}$): $90$ bits, $95$ bits, and $100$ bits.
The reward measures the system’s total transmission rate, achieved by maintaining both user-specific minimum transmission rate constraints and the minimum MI constraint.
From the results, we can observe that the reward increases with iterations, stabilizing into a steady state after approximately $1000$ iterations.
A peak reward of approximately $11.8 \times 10^{12}$ bps is achieved with the least restrictive $\mbox{MI}^{\min}$ of $90$ bits.  
As the $\mbox{MI}^{\min}$ increases to $95$ bits, the reward plateaus at about $11.2 \times 10^{12}$ bps.
With the most stringent requirement of $100$ bits, the reward further decreases, settling at around $10.5 \times 10^{12}$ bps. 
These outcomes illustrate a clear trade-off: lower $\mbox{MI}^{\min}$ values facilitate higher transmission rates, whereas higher values, enforcing stricter constraints, cap the achievable rates.

\begin{figure}[!t]
\centering
\includegraphics[width=\linewidth]{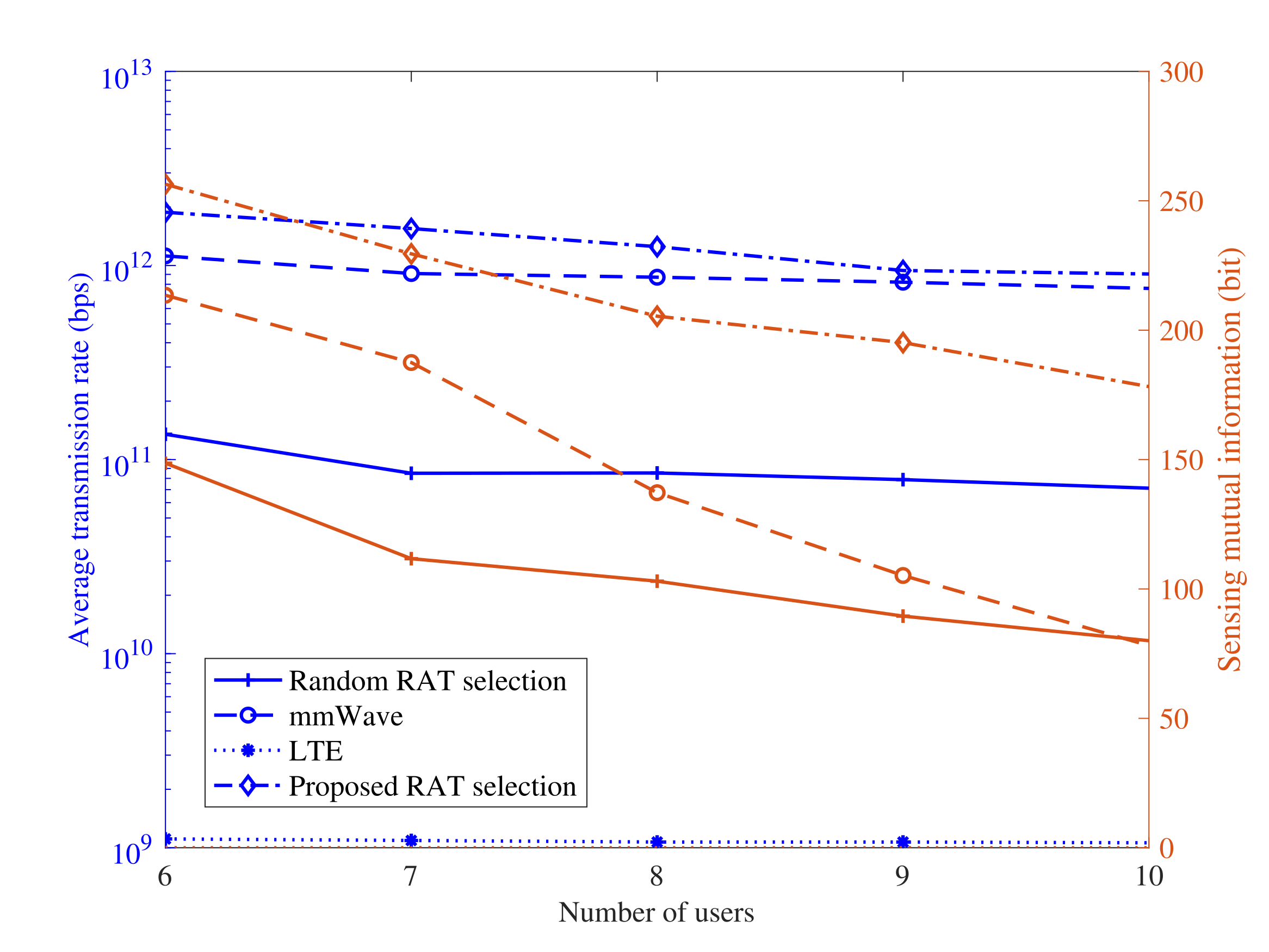}
\caption{The average transmission rate and sensing mutual information versus the increase of number of users.}
\label{fig:4}
\end{figure}        

Figure \ref{fig:4} compares the performance in terms of average transmission rate (left $y\mbox{-axis}$) and sensing MI (right $y\mbox{-axis}$) for all the mentioned methods as the number of users increases from $6$ to $10$. 
The results clearly show that our proposed RAT selection method consistently surpasses others, ranking in performance as follows: Proposed RAT selection $\succ$ mmWave $\succ$ Random RAT selection $\succ$ LTE. 
It not only delivers the highest average transmission rate but also achieves the greatest sensing MI across all scenarios. 
Even with an increasing number of users, it demonstrates only a moderate decline in performance, effectively maintaining a strong equilibrium between transmission rate and sensing MI. These outcomes highlight the exceptional ability of our method to provide reliable transmissions and maintain sufficient MI, essential for precise activity recognition.

\section{Conclusion}
In this paper, we proposed a novel CV-aided ISAC framework that addresses the challenge of maintaining reliable communication for users obstructed from direct mmWave communication.
By leveraging camera-assisted user localization, the framework enables dynamic RAT selection, allowing the BS to switch between mmWave and LTE communication depending on user visibility.
Additionally, the SlowFast model was integrated to analyze user behavior from camera-captured images, ensuring  users' minimum transmission rates are aligned with their activities.
Our research focused on maximizing the total transmission rate for all users while ensuring that sensing MI and minimum transmission rate constraints are met.
To achieve this, we developed a DDPG-based algorithm, which optimizes the precoding  and RAT selection, ensuring efficient resource allocation under varying environmental conditions.
The effectiveness of the proposed framework and the optimization strategy was validated through numerical simulations, demonstrating the practical viability of the approach in ensuring continuous, high-quality communication while meeting the sensing requirements.

%


\bibliographystyle{IEEEbib}
\bibliography{refs}

\end{document}